\documentclass[prl,aps,twocolumn,amsmath,amssymb,showpacs]{revtex4}

\usepackage{dcolumn}
\usepackage{bm}
\usepackage{graphicx}
\usepackage{psfrag}
\usepackage{subfigure}

\begin{document}
\title{Magnetic properties of spin-orbital polarons in lightly doped cobaltates}
\author {M. Daghofer, P. Horsch, and G. Khaliullin}
\affiliation{ Max-Planck-Institut f\"ur Festk\"orperforschung,
              Heisenbergstrasse 1, D-70569 Stuttgart, Germany }
\date{May 12, 2006}

\begin{abstract}
We present a numerical treatment of a spin-orbital polaron model for
Na$_x$CoO$_2$ at small hole concentration ($0.7 < x< 1$). We
demonstrate how the polarons account 
for  the peculiar magnetic properties of
this layered compound: They explain the large susceptibility; 
their internal degrees of freedom lead 
both to a negative Curie-Weiss temperature and yet to a ferromagnetic
intra-layer interaction, thereby resolving a puzzling contradiction between
these observations. We make specific predictions on the momentum and energy location of
excitations resulting from the internal degrees of freedom of the polaron, and
discuss their impact on spin-wave damping. 
\end{abstract}


\maketitle

The complex physics of transition metal oxides often leads to intriguing and
potentially rewarding properties like, e.g., high-$T_C$ superconductivity in
the case of cuprates or colossal magnetorestistance in the case of
manganites. The layered cobaltate Na$_x$CoO$_2$ is of interest mainly for
two reasons: When it is hydrated, the resulting Na$_x$CoO$_2$:$y$H$_2$O
becomes super-conducting for doping $x\approx 0.3, y\approx1.4$
\cite{Tak03}. Without hydration, it shows an exceptionally high thermopower,
i.e. the capability to transform temperature differences into electricity, for
$0.5\leq x\leq 0.9$ \cite{Fuj01,Mik03} in an unusual combination 
with low resistivity. Magnetic fields strongly affect the
thermopower in this material \cite{Wan03}, which suggests that
electron-electron correlations and spin degrees of freedom are
important. These remarkable features of cobaltates have triggered large
interest due to their potential applications, therefore, the origin of the
exotic transport and magnetic properties of Na$_x$CoO$_2$ is one of the hot
topics in the field of strongly correlated materials.

While Na$_x$CoO$_2$ does not show magnetic order either for $x< 0.7$ (except
at $x=0.5$)
\cite{Foo04} or for $x=1$ \cite{Lan05}, it is found in the $A$-type
antiferromagnetic phase below $T_N\approx 20$K for $0.7 < x < 0.9$
\cite{Bay05,Hel05}. In this phase, the carriers move on a hexagonal lattice within 
the ferromagnetic planes, which are antiferromagnetically stacked. The lattice
spacing within the planes is much 
smaller than between them which typically leads to considerably stronger
in-plane coupling in such layered systems, e.g. in NaNiO$_2$ with the same crystal
structure \cite{Lew05}. Surprisingly, this is not the case here: The
ferromagnetic in-plane coupling is only of the order of magnitude of the
antiferromagnetic inter-layer coupling \cite{Bay05} or even smaller
\cite{Hel05}.
At higher temperatures $T\gg T_N$, the susceptibility
$\chi$ shows Curie-Weiss behavior, i.e. $\chi\propto 1/(T-\theta)$, with a
\emph{negative} Curie-Weiss temperature $\theta < 0$ \cite{Mik03,Gav04,
  Wan03,Mot03, Bay04}. As has been pointed out in Ref.~\onlinecite{Bay05}, this
contradicts the \emph{positive} Curie-Weiss temperature inferred from
spin-wave data.  

In this letter, we analyze 
a spin-orbital polaron model proposed in Ref.~\onlinecite{Kha05} for Na$_x$CoO$_2$ at large $x$ (i.e. small hole
concentration). By means of unbiased numerical techniques, we calculate static
and dynamic observables and thereby provide the first quantitative description of
the exotic magnetic properties including the negative Curie-Weiss temperature
$\theta$, the magnitude and temperature dependence of the susceptibility, as
well as
a small ferromagnetic in-plane coupling mediated by dilute polarons. Further,
we discuss the expected impact on magnetic response and predict, apart from
the observed spin-waves, additional scattering at specific momenta and energies, which can
be verified experimentally. 

\begin{figure}
  \psfrag{Jd}{$J_\textrm{diag}$}
  \psfrag{J}{$J$}
  \psfrag{J'}{$J'$}
  \psfrag{a}{(a)}
  \psfrag{b}{(b)}
  \psfrag{c}{(c)}
  \psfrag{0}{$S=0$}
  \psfrag{1}{$S=1$}
  \psfrag{e}{$e_g$}
  \psfrag{t}{$t_{2g}$}
  \psfrag{Jij}{$J_{ij}$}
  \psfrag{Jh}{\small$j_H$}
  \vspace*{-3em}
  \subfigure{\label{fig:orbs}}
  \subfigure{\label{fig:SE}}
  \subfigure{
  \includegraphics[width=0.47\textwidth]{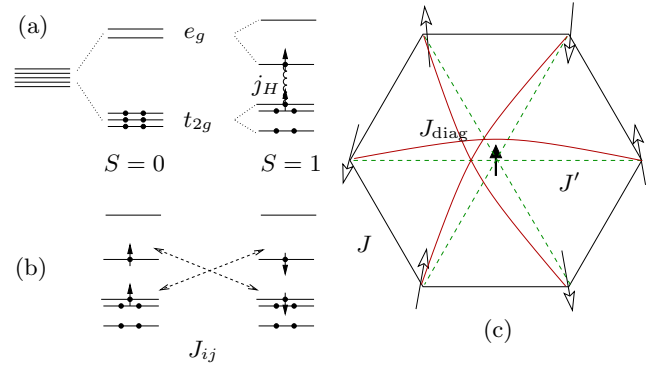}\label{fig:polaron}}
  \vspace*{-1em}
  \caption{(Color online)
    (a) Cubic symmetry for the Co$^{3+}$-ion is removed next to a hole
    (Co$^{4+}$-ion), this reduces the energy splitting between $t_{2g}$ and
    $e_g$ levels of Co$^{3+}$; Hund's rule coupling $j_H$ then stabilizes
    $S=1$ instead of $S=0$.     
    (b) Virtual $t_{2g}$-$e_g$ hopping leading to antiferromagnetic coupling $J_{ij}$
    between two Co$^{3+}$ ions.  
    (c) Spin structure and interactions within the polaron: Co$^{4+}$ with $s=1/2$ in the center, induced spin
    $S=1$ on the six adjacent Co$^{3+}$-sites.  (Co$^{3+}$-ions
    further away have $S=0$.) Parameters: $J$ couples nearest neighbor (n.n.) spins with
    $S=1$ on the ring, $J_\textrm{diag}$ couples them across the diagonal, both
    are antiferromagnetic. $J'$ couples the central $s=1/2$ to the
    ring. \label{fig:orbs_polaron}} 
\end{figure}

In NaCoO$_2$, the Co-$d$ orbitals are split into energetically lower $t_{2g}$
and higher $e_g$ states, and the  six electrons of the Co$^{3+}$-ions occupy
the $t_{2g}$ orbitals, leading to a non-magnetic $t_{2g}^{6}$
configuration. Consequently, NaCoO$_2$ has rather small susceptibility
\cite{Lan05}. Removing some of the sodium induces holes and the
Co$^{4+}$-sites are found in a $t_{2g}^{5}$ configuration with spin
$s=1/2$. The few dilute spins embedded in  a non-magnetic background,  
however, would give orders of magnitude smaller susceptibility at room
temperature than observed \cite{Mik03, Mot03}. The key to the problem is found in the
orbital degree of freedom: The Co$^{4+}$-holes removes cubic symmetry at the 
six surrounding Co$^{3+}$-ions, this splits the $t_{2g}$-triplet and the
$e_g$-doublet (on top of the trigonal splitting induced by the layered crystal
structure) and thus reduces the energy gap between the highest $t_{2g}$-
and the lower $e_g$-orbitals, see
the schematic representation in Fig.~\ref{fig:orbs}. The gap actually
becomes smaller than Hund's 
rule splitting, i.e., energy can be gained by transferring one of the
electrons from the highest $t_{2g}$-level into the lowest $e_g$-orbital and
forming a triplet with the remaining unpaired electron \cite{Ber04,Kha05}.

The mechanism of polaron formation by holes inducing an energy splitting
between orbitals is similar to the orbital-polaron formation considered for
lightly doped LaMnO$_3$ \cite{Kil99}, however, the 90-degree angle of the Co-O-Co bonds
causes an important difference: While the spin coupling is
ferromagnetic in manganites (leading to a large total spin of the polaron),
the coupling in the present case is \emph{anti}ferromagnetic \cite{Kha05}.
Because of the 90$^\circ$ angle, $e_g$-$e_g$-exchange is suppressed, and the
dominant $t_{2g}$-$e_g$ process, shown in Fig.~\ref{fig:SE}, favors
antiferromagnetic alignment of the two ions, as well as an analogous
$t_{2g}$-$t_{2g}$ hopping. The resulting spin-orbital polaron is depicted in
Fig.~\ref{fig:polaron}. A similar  antiferromagnetic exchange
$J_\textrm{diag}$, induced via virtual longer-range hopping $t'$, couples the
$S=1$-sites along the diagonals of the polaron and they also couple to the central
$s=1/2$. These dressed carriers instead of bare holes are in agreement
with ARPES measurement indicating a strongly renormalized quasi-particle band
width \cite{Has04,Yan04}. 

The effective spin Hamiltonian resulting from these considerations is given by:
\begin{equation}
H = J \sum_{i = 1}^6 {\vec S}_i {\vec S}_{i+1} 
+ J_\textrm{diag}\sum_{i = 1}^3 {\vec S}_i {\vec S}_{i+3}
+ J'\sum_{i = 1}^6 {\vec s}_0 {\vec S}_{i}\;,
\end{equation}
where $S_i, i=1,\dots,7, S_7\equiv S_1$ denote the $S=1$ spins in the outside
ring and $s_0$ the $s=1/2$ in the center. $J$, $J_\textrm{diag}$, and $J'$ are
the coupling constants: From the orbital structure,
$J\approx 10-20\,\textrm{meV}, J_\textrm{diag}\lesssim J$ can be inferred \cite{Kha05}. The
coupling $J'$ of the central $s=1/2$ to the outside spins is less clear
because ferromagnetic and antiferromagnetic contributions compete, it can {\it
  a priori} be positive or negative. As the Hilbert space of this Hamiltonian
is only 1458, we easily diagonalize it and compute observables at any temperature.

\begin{figure}
  \vspace*{-2em}
  \subfigure{\label{fig:inv_susc}}
  \subfigure{
  \includegraphics[width=0.44\textwidth]{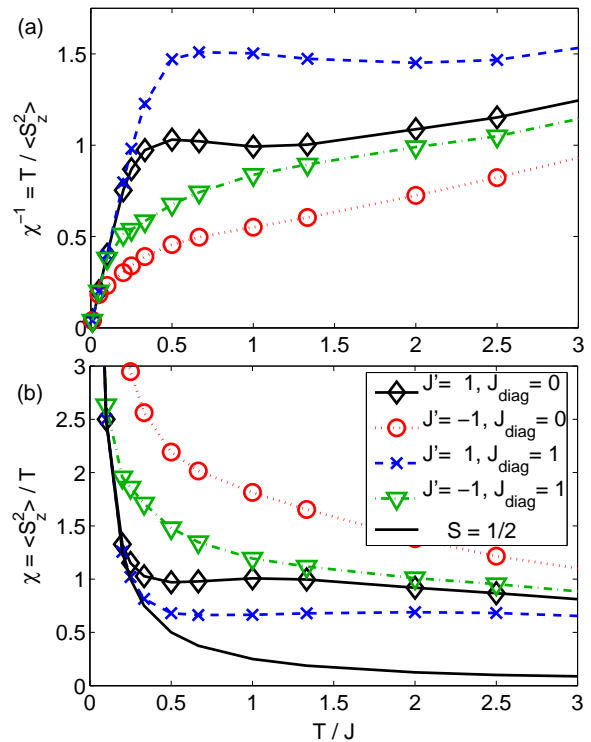}\label{fig:susc}}\vspace*{-1em}
  \caption{(Color online) (a) Inverse susceptibility and (b) susceptibility of a polaron, see Fig.~\ref{fig:polaron}, for some
    parameter values. For comparison, the susceptibility resulting from a
    single $s=1/2$ spin is included  in(b) (solid black line without symbols).\label{fig:susc_inv_susc}}
\end{figure}

The susceptibility is particularly interesting:
\begin{equation}
\chi(T) = \langle (S_\textrm{tot}^z)^2 \rangle / T 
     = \sum_l \frac{\textrm{e}^{-\frac{E_l}{k_bT}}}{ZT} \langle E_l|\left(\sum_{i=0}^6 S_i^z\right)^2 |E_l\rangle\;,
\end{equation}
with $S_\textrm{tot}^z$ the $z$ component of the total spin and
temperature $T$. $E_l$ gives the eigenenergy of state $|E_l\rangle$, $S_i^z$
the $z$-component of the spin at site $i$, $k_b$ denotes the Boltzmann
constant and $Z=\sum_l \textrm{e}^{-\frac{E_l}{k_bT}}$ the partition function.
The inverse susceptibility $\chi^{-1}$ of a polaron is depicted in
Fig.~\ref{fig:inv_susc} for several parameter sets. 
For all cases, the slope increases for decreasing $T$, that is,
$\langle (S_\textrm{tot}^z)^2 \rangle$ becomes smaller.
The reason is the freezing of the $S=1$ system into a singlet state as the
antiferromagnetic bonds become stronger at $T\ll J$,  which leaves only the
net $s=1/2$ spin. If the high-temperature part were continued as a straight 
line, it would cross the $T$-axis at $\theta < 0$, i.e. give a negative
Curie-Weiss temperature. 

Figure~\ref{fig:susc} shows the susceptibility of a polaron $\chi(T)$
depending on temperature for the same parameter sets. We note
that comparing the shapes of the curves to experimental susceptibility
data  \cite{Bay04, Foo04, Mik03} points towards $J'<0$ . 
For comparison, we also included the susceptibility $\chi = 1/4T$ of a
single $s=1/2$, which would be obtained, if we had only the $s=1/2$
resulting from the Co$^{4+}$-site, i.e., 
if all Co$^{3+}$-ions would remain in  the $S=0$ state: For $T\gtrsim J$, it
is much smaller than the susceptibility of the polaron,
regardless of parameter values. Experimentally, the high-temperature values
for the susceptibility $\chi_\textrm{exp}$ for $0.7 \lesssim x\lesssim 0.9$ range from   
$\sim 10^{-4}\textrm{emu/mol}$ to  $\sim 10^{-3}
\textrm{emu/mol}$ \cite{Mik03, Mot03, Sug04, Bay04, Foo04, Lan05}. From our high-temperature values
for the susceptibility of a single polaron $\chi_\textrm{polaron} \approx
1/J$, see Fig.~\ref{fig:susc}, the susceptibility per mole is obtained via 
\begin{equation}
\label{eq:susc_theor}
\chi_\textrm{theory} = \frac{g^2 \mu_B^2 N_A}{J} (1-x)\, \chi_\textrm{polaron}\;,
\end{equation}
with Land{\'e} g-factor $g=2$, Bohr magneton $\mu_B$, Avogadro's constant
$N_A$ and doping $x$. Taking $J=20\, \textrm{meV}, x = 0.82$, this corresponds
to $\chi_\textrm{theory}\approx 1.2 \times 10^{-3}$, in accordance with the
value reported for Na$_{0.82}$CoO$_2$ \cite{Bay04}. Both the shape, see
Fig.~\ref{fig:susc} and the absolute values of the susceptibility obtained
with the polaron model therefore agree well with experiment.

\begin{figure}
  \psfrag{Jd}{$J_\textrm{diag}$}
  \psfrag{Jab}{$J_\textrm{ab}$}
  \psfrag{J}{$J$}
  \psfrag{J'}{$J'$}
  \psfrag{a}{(a)}
  \psfrag{b}{(b)}
  \psfrag{c}{(c)}
  \psfrag{s_a}{$s_a$}
  \psfrag{s_b}{$s_b$}
  \vspace*{-3.5em}
  \subfigure{\label{fig:bip}}
  \subfigure{\label{fig:dbl_s}}
  \subfigure{
    \includegraphics[width=0.47\textwidth]{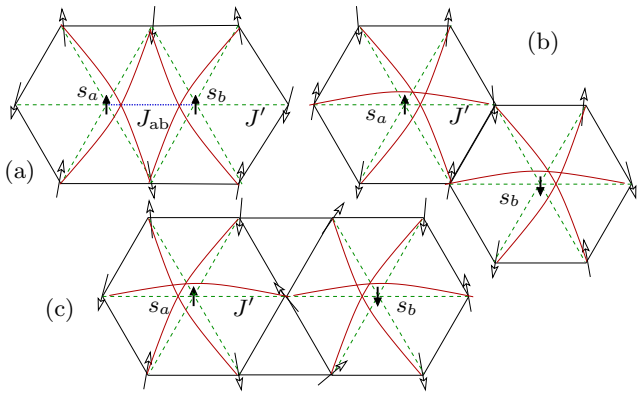}\label{fig:dbl_e}}
  \caption{(Color online) Three possible configurations of interacting polarons. $J_{ab}$ gives
    the bare ferromagnetic coupling between $s_a$ and $s_b$, remaining couplings and symbols as
    Fig.~\ref{fig:polaron}.\label{fig:dbl}} 
\end{figure}

\begin{figure}
  \psfrag{Jd}{$J_\textrm{diag}$}
  \psfrag{Jab}{$J_\textrm{ab}$}
  \psfrag{J}{$J$}
  \psfrag{J'}{$J'$}
  \psfrag{s_a}{$s_a$}
  \psfrag{s_b}{$s_b$}
\includegraphics[width=0.44\textwidth]{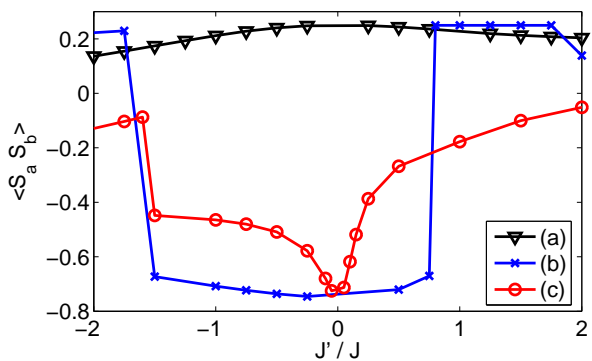}
  \caption{(Color online) Ground-state spin correlations $\langle \vec
    s_a\vec s_b \rangle$ between the two Co$^{4+}$ sites for the
    interacting-polaron configurations depending on $J'$. (a), (b) and (c)
    indicate the cluster as depicted in Fig.~\ref{fig:dbl}. $J_\textrm{diag} =
    J, J_{ab} = 0$.\label{fig:SaSb_bipolaron}}  
\end{figure}

\begin{figure}
\includegraphics[width=0.4\textwidth]{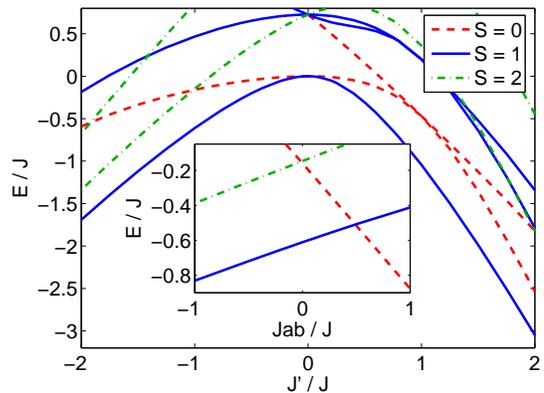}
\caption{(Color online) Energies of the lowest states with $S_\textrm{tot}^z = 0, 1,2$ for the
  bi-polaron Fig.~\ref{fig:bip} depending on $J'$ for $J_{ab}=0$ (main
  panel) and depending on $J_{ab}$ for $J'=-J$(inset). $J_\textrm{diag} =
  J$.\label{fig:levels_bipolaron}}  
\end{figure}

While we have solved the riddle of the susceptibility, we have still to show how
ferromagnetic interactions within the plane can arise and lead to the
$A$-type antiferromagnetism observed below 20K \cite{Bay05, Hel05}. To this end, we
analyze three configurations of interacting polarons depicted in
Fig.~\ref{fig:dbl}, which will
occasionally occur as two mobile polarons come near each other. At first, we examine the direct 
coupling $J_{ab}$ between two Co$^{4+}$-ions situated at n.n. sites as in the
bi-polaron Fig.~\ref{fig:bip}. Like the coupling between
Co$^{3+}$-spins $J$, it originates from virtual hopping processes over a
90-degree Co-O-Co path; it is composed of both
ferromagnetic and antiferromagnetic terms. With an average hole density
$n=1/3$ in each $t_{2g}$ orbital, we can obtain it from
Eqs. (5.5.) to (5.7.) of Ref.~\onlinecite{Kha05}, with $A\approx B\approx C/2$. We arrive at
$J_{ab}=-8/9A$, which is $\sim -J$, because $A$ and $J$ are of the same
scale. The ferromagnetic net interaction is due to a dominance of 
Hund's-rule coupling over the other terms.

However, it is not obvious that the more complex polarons do not mediate a
different and possibly antiferromagnetic interaction and thus turn the net
in-plane interaction from ferro- to antiferromagnetic. We therefore solve the
complete model given by the clusters in Fig.~\ref{fig:dbl}, 
and as the Hilbert space is
here larger, we employ the Lanczos algorithm. From the low-energy eigenstates,
we calculate observables like the spin-spin correlation between the two Co$^{4+}$-sites  $\langle \vec 
s_a\vec s_b \rangle$, which we show in Fig.~\ref{fig:SaSb_bipolaron} in absence of $J_{ab}$. For the bi-polaron
Fig.~\ref{fig:bip}, we see that the ground state
favors ferromagnetic alignment, while it is
antiferromagnetic for the two clusters with larger distance between the
Co$^{4+}$-ions.

In order to make a statement on the size of the resulting effective exchange,
we have to consider the involved energies. For the bi-polaron
Fig.~\ref{fig:bip}, i.e. for the n.n. case, they are shown in
Fig.~\ref{fig:levels_bipolaron} for different values of the 
$z$-component of the total spin (a good quantum number)
$S_\textrm{tot}^z=0,1,2$. As expected, the singlet and triplet states are
degenerate for  $J'=0$, i.e., when the Co$^{4+}$ states are not coupled to the
Co$^{3+}$ system. For $|J'|>0$, the ferromagnetic triplet state has lower
energy than the antiferromagnetic singlet state, i.e., the effective
interaction via the $S=1$-system strengthens the direct ferromagnetic interaction. 
The effective interaction is then combined with the direct coupling $J_{ab}$ and  the inset of
Fig.~\ref{fig:levels_bipolaron} shows the energies of the lowest bi-polaron
states for $J'=-J$ depending on $J_{ab}$. We see that the energy change is
$\approx J_{ab}\langle s_a s_b\rangle$, i.e. the two 
contributions can be simply added $\tilde J_\textrm{bipol.} \approx
J_\textrm{bipol.} + J_{ab}$.

\begin{figure}
  \includegraphics[width=0.4\textwidth]{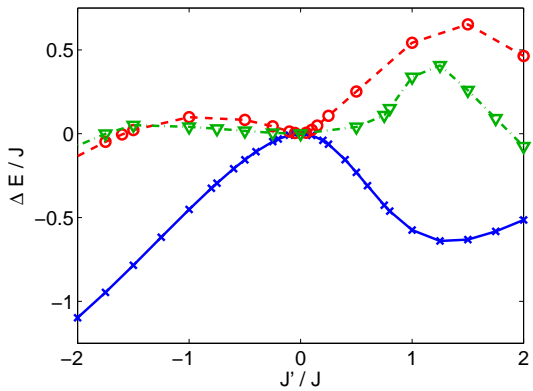}
  \caption{(Color online) Singlet-triplet gap depending on $J'$ for all
    three clusters: crosses for the bi-polaron
    Fig.~\ref{fig:bip}, triangles for Fig.~\ref{fig:dbl_s}, and
    circles for Fig.~\ref{fig:dbl_e}. $J_\textrm{diag} = J$
    \label{fig:delta_s_t}}    
\end{figure}

Next, we perform a similar analysis on the two other configurations of edge-
or corner-sharing polarons. As can be seen in Fig.~\ref{fig:SaSb_bipolaron}, we actually
find that the ground states have antiferromagnetically coupled Co$^{4+}$.
The energy difference between the  lowest triplet state and the singlet ground
state, which is equivalent to an
effective \emph{anti}ferromagnetic next nearest neighbor (n.n.n.) coupling, is depicted in
Fig.~\ref{fig:delta_s_t}. While it nearly vanishes for $J'<0$, it becomes
comparable to the ferromagnetic n.n. exchange $J_{ab}$ for $J'>0$, especially if
one considers that there are twice as many possible n.n.n. interactions as
n.n. ones (i.e. 12 instead of 6). Such an effective  antiferromagnetic
long-range coupling between the polarons should manifest itself as a deviation
from a n.n. Heisenberg-like dispersion in spin-wave measurements. While the fact
that such a deviation was not observed might result from too small momentum
vectors in experiment, it could also indicate that $J'$  is, in fact, negative,
which leads to negligible n.n.n.-coupling, see Fig.~\ref{fig:delta_s_t}. The
spin-wave dispersion thereby provides a way to experimentally determine the
sign of $J'$. 

For the n.n. coupling and $|J'|\lesssim J$, which we believe to be a realistic
range based on the susceptibility data, we read off a ferromagnetic exchange
$0>J_\textrm{bipol.} \gtrsim -0.5 J$ (see Fig.~\ref{fig:delta_s_t}) in
addition to the direct $J_{ab}\sim -J$. However, the ferromagnetic coupling is
substantially reduced by the fact 
that only some bonds are covered by bi-polarons. For hole density $\delta \equiv
1-x$, we have $ J_\textrm{av} \simeq  (J_\textrm{bipol.} + J_{ab}) \cdot \delta $.
At $\delta\approx 0.2$ and $J'=-J$, this gives an average exchange $J_\textrm{av}S_iS_j$
with $J_\textrm{av} \sim -6$~meV, which agrees well with measured values
of $\sim -8$~meV \cite{Bay05, note} and $\sim -6$~meV \cite{Hel05}.

We have already shown that the internal degrees of freedom of the polaron, i.e. the energy-level
structure of the $s=1/2$ coupled to a $S=1$ ring with a Haldane gap, are
crucial for the high-temperature behavior of the static 
susceptibility $\chi$. They likewise influence the dynamic susceptibility
$\chi''(\vec q, \omega)$, where they lead to a quasi-localized mode
\cite{Kha05} in addition to the collective and dispersive spin waves. 
For its experimental observation, it is crucial to know energy and momentum of
this mode. We predict here that its largest weight in momentum 
space is found at $\vec q=K = 
(4\pi/3,0,0)$, a little is also obtained for $\vec q=M= (0,
2\pi/\sqrt{3},0)$, but the weight is quickly reduced upon approaching the
$\Gamma$-point $(0,0,0)$ (details for the  structure factor will be given
elsewhere). The energy of the lowest peak is found at $\omega \sim J$ for 
$J_\textrm{diag} = J, J'=0$, and is lowered by a coupling $|J'|>0$ to the
central $s=1/2$.  In spin-wave experiments, large damping is expected at these
energies and momenta because  the spin waves strongly scatter on the
quasi-localized modes.
For ferromagnetic $J' < 0$, which has already been noted
before to be more plausible based on susceptibility and spin-wave data, damping is expected
to start between $\omega \sim J/2$ and $\omega \sim J$, depending on $J'$, as actually observed in
neutron-scattering experiments \cite{Bay05,Hel05}. 
For an antiferromagnetic $J'>0$, the energy of the localized mode is reduced  much
faster and reaches $\omega=0$ for $J'=J$. This seems to rule out large antiferromagnetic
$J'$, because spin-waves are actually clearly measurable at not too large
energies \cite{Bay05,Hel05}.

To summarize, we present a coherent and quantitative description of the
magnetic properties of Na$_x$CoO$_2$ at large $x$. 
The concept of spin-orbital polarons could likewise be a 
key in understanding the unconventional transport properties of cobalt-based
materials. 

\begin{acknowledgments}
We would like to thank B. Keimer, C. Bernhard, S. Bayrakci and A. T. Boothroyd for
useful communications. 
\end{acknowledgments}


\begin{thebibliography}{10}

\bibitem{Tak03}
{K. Takada} {\it et~al.}, Nature (London) {\bf 422},  53  (2003).

\bibitem{Fuj01}
{K. Fujita}, {T. Mochida}, and {K. Nakamura}, Jpn. J. Appl. Phys. {\bf 40},
  4644  (2001).

\bibitem{Mik03}
{M. Mikami} {\it et~al.}, 7383 {\bf 42},  Jpn. J. Appl. Phys.  (2003).

\bibitem{Wan03}
{Y. Wang}, {N. S. Rogado}, {R. J. Cava}, and {N. P. Ong}, Nature (London) {\bf
  423},  425  (2003).

\bibitem{Foo04}
{M. L. Foo} {\it et~al.}, Phys. Rev. Lett. {\bf 92},  247001  (2004).

\bibitem{Lan05}
{G. Lang} {\it et~al.}, Phys. Rev. B {\bf 72},  094404  (2005).

\bibitem{Bay05}
{S. P. Bayrakci} {\it et~al.}, Phys. Rev. Lett. {\bf 94},  157205  (2005).

\bibitem{Hel05}
{L. M. Helme} {\it et~al.}, Phys. Rev. Lett. {\bf 94},  157206  (2005).

\bibitem{Lew05}
{M. J. Lewis} {\it et~al.}, Phys. Rev. B {\bf 72},  014408  (2005).

\bibitem{Gav04}
{J. L. Gavilano} {\it et~al.}, Phys. Rev. B {\bf 69},  100404(R)  (2004).

\bibitem{Mot03}
{T. Motohashi} {\it et~al.}, Phys. Rev. B {\bf 67},  064406  (2003).

\bibitem{Bay04}
{S. P. Bayrakci} {\it et~al.}, Phys. Rev. B {\bf 69},  100410(R)  (2004).

\bibitem{Kha05}
G. Khaliullin, Prog. Theor. Phys. Suppl. {\bf 160},  155  (2005).

\bibitem{Ber04}
{C. Bernhard} {\it et~al.}, Phys. Rev. Lett. {\bf 93},  167003  (2004).

\bibitem{Kil99}
{R. Kilian} and {G. Khaliullin}, Phys. Rev. B {\bf 60},  13458  (1999).

\bibitem{Has04}
{M. Z. Hasan} {\it et~al.}, Phys. Rev. Lett. {\bf 92},  246402  (2004).

\bibitem{Yan04}
{H.-B. Yang} {\it et~al.}, Phys. Rev. Lett. {\bf 92},  246403  (2004).

\bibitem{Sug04}
{J. Sugiyama} {\it et~al.}, Phys. Rev. B {\bf 69},  214423  (2004).

\bibitem{note}
\textnormal{{Note} that our definition of {$J$} differs by a factor of 2 from
  the convention {$2JS_iS_j$} used in \cite{Bay05}}.

\end{thebibliography}
\end{document}